\documentclass[twocolumn, showpacs,preprintnumbers,amssymb]{revtex4}
\usepackage[dvips]{color}
\usepackage{epsfig}

\def\NPB{{Nucl. Phys.} { B}}
\def\PLB{{Phys. Lett.} B}
\def\PRL{Phys. Rev. Lett.}
\def\PRD{{Phys. Rev.} D}
\def\Vec#1{\mbox{\boldmath $#1$}}
\begin{document}

\title{Domain Wall Fermion Lattice Simulation\\ in Quaternion Basis}

%\classification{12.38.-t, 12.38.Gc, 12.60.-i, 11.10.Lm}
%\keywords      {Quaternion, Octonion, Triality, Proton charge form factor}

\author{Sadataka Furui}
\email{furui@umb.teikyo-u.ac.jp}
\affiliation{School of Science and Engineering, Teikyo University\\
 1-1 Toyosatodai, Utsunomiya 320-8551,Japan }

%\author{<author2>}{
%  address={<common address for author2 and author3>}
%}

%\author{<author3>}{
%  address={<common address for author2 and author3>}
%  ,altaddress={<author1 address>} % additional visiting address
%}

\begin{abstract}
 In the QCD analysis, when quarks are expressed in quaternion basis, the quark and its charge conjugate together are expressed by octonions and the octonion posesses the triality symmetry. Gluos are expressed by Pl\"ucker coordinates of spinors. Roles of triality in the proton charge form factor, three loop gluon self energy, technicolor, fine tuning and unparticle physics are discussed. 

\end{abstract}
\pacs{12.38.-t, 12.38.Gc, 12.60.-i, 11.10.Lm}
\maketitle

%%%%%%%%%%%%%%%%%%%%%%%%%%%%%%%%%%%%%%%%%%%%
%% MAINMATTER
%%%%%%%%%%%%%%%%%%%%%%%%%%%%%%%%%%%%%%%%%%%%

\section{Introduction} 
Domain wall fermion(DWF) approximately preserves chiral symmetry and it transforms under $SU(3)$ color and $SU(2)$ spin as symmetries of internal coordinates. Although Pauli matrices which follows the SU(2) symmetry is frequently used, the symmetry of quaternion ${\mathcal H}$ which is invented by Hamilton is not considered seriously.
By adding a new imaginary unit ${ l}$ orthogonal to the quaternion basis ${\bf e_1}={ i},{\bf e_2}={j},{\bf e_3}={ k}$,  one can construct octonion 
 ${\mathcal O} ={ H}+{l H}$ which is spanned by
\begin{equation}
\{ 1,{\bf i},{\bf j},{\bf k},{\bf l},{\bf i}{\bf l},{\bf j}{\bf l},{\bf k}{\bf l} \}=\{1,{\bf e_1, e_2, e_3, e_4, e_5, e_6, e_7}\}\nonumber
\end{equation}
i.e. one real unit and 7 imaginary units \cite{Lo01}.

In this 8 dimensional space, \'E. Cartan introduced a universal covering group of $SO(8)$, which is called $Spin(8)$. It has the triality automorphism.  

 In this presentation, I show that the triality automorphism could be an important ingredient which can solve various puzzles in the infrared (IR) QCD. In sect.2, I introduce the quaternion, the octonion and triality automorphism and in sect.3 puzzles in IR QCD are discussed. The lattice simulation of proton charge form factor using quaternion bases, assuming correlation of domain wall fermions via exchange of self-dual gauge fields \cite{SF10} is shown in sect.4. Discussion and conclusion are given in sect.5. 

\section{Quaternion, Octonion and triality}
In 1877 Frobenius showed that an associative, quadratic real algebra ${\mathcal A}$ without divisors of zero has only three possibilities
\begin{enumerate}
\item ${\mathcal A}$ is isomorphic to ${\mathcal R}$ (Real).
\item ${\mathcal A}$ is isomorphic to ${\mathcal C}$ (Complex).
\item ${\mathcal A}$ is isomorphic to ${\mathcal H}$ (Quaternion).
\end{enumerate}

Quaternions are generalization of complex number ${\mathcal C}={\mathcal R}+i{\mathcal R}$, which are expressed as $q=w+x{\Vec i}+y{\Vec j}+z{\Vec k}$. Automorphism group of ${\mathcal H}={\mathcal R}+{\mathcal R}^3$ is SO(3).

A new imaginary unit $l$ that anticommutes with the bases of quaternions ${\Vec i},{\Vec j},{\Vec k}$ compose octonions ${\mathcal O}={\mathcal H}+l{\mathcal H}$. Automorphism group of ${\mathcal O}={\mathcal R}+{\mathcal R}^7$ is not SO(7), but exceptional Lie group $G_2$. It contains tensor product of three ${\mathcal R}^7$ bases and three vectors.
The triality automorphism is a transformation that rotates 24 dimensional bases defined by Cartan\cite{Cartan66}.

\begin{eqnarray}
&&\{\xi_0, \xi_1, \xi_2, \xi_3, \xi_4\},\quad
\{\xi_{12}, \xi_{31}, \xi_{23}, \xi_{14}, \xi_{24}, \xi_{34}\},\nonumber\\
&&
\{\xi_{123}, \xi_{124}, \xi_{314}, \xi_{234}, \xi_{1234}\},\nonumber\\
&&
\{x^1, x^2, x^3, x^4\}, \quad \{x^{1'}, x^{2'}, x^{3'}, x^{4'}\}
\end{eqnarray}

There are three semi-spinors which have a quadratic form which is invariant with respect to the group of rotation
\begin{eqnarray}
&&\Phi={^t\phi} C\phi=\xi_0\xi_{1234}-\xi_{23}\xi_{14}-\xi_{31}\xi_{24}-\xi_{12}\xi_{34}\nonumber\\
&&\Psi={^t\psi} C\psi=-\xi_1\xi_{234}-\xi_{2}\xi_{314}-\xi_{3}\xi_{124}+\xi_{4}\xi_{123}\nonumber\\
\end{eqnarray}
and the vector
\begin{equation}
F=x^1 x^{1'}+x^2 x^{2'}+x^3 x^{3'}+x^4 x^{4'}
\end{equation}

With use of the quaternion bases $1,{\Vec i},{\Vec j}, {\Vec k}$, the spinors $\phi$ and $C\phi=\phi'$ are defined as
\begin{eqnarray}
\phi&=&\xi_0+\xi_{14}{\Vec i}+\xi_{24}{\Vec j}+\xi_{34}{\Vec k}\nonumber\\
C\phi&=&\xi_{1234}-\xi_{23}{\Vec i}-\xi_{31}{\Vec j}-\xi_{12}{\Vec k}.
\end{eqnarray}
Similarly, $\psi$ and $C\psi=\psi'$ are defined as
\begin{eqnarray}
\psi&=&\xi_4+\xi_1{\Vec i}+\xi_2{\Vec j}+\xi_3{\Vec k}\nonumber\\
C\psi&=&\xi_{123}-\xi_{234}{\Vec i}-\xi_{314}{\Vec j}-\xi_{124}{\Vec k}.
\end{eqnarray}

\section{Correlation of quarks via self-dual gauge field}
\subsection{Proton charge form factor}
To calculate proton charge form factor with use of $16^3\times 32\times 16$ DWF produced by RBC/UKQCD collaboration \cite{DWF}, I first perform Landau gauge fixing and then Coulomb gauge fixing of the gauge configuration.  Instead of performing the residual gauge transformation of the Coulomb gauge, I rotate the fermion on the left domain wall and on the right domain wall such that they are correlated by the self dual gauge field which is parametrized as Corrigan and Goddard\cite{CG81}.   

 The transition function of \cite{CG81} is
\[
g(\lambda\omega,\lambda\pi)=g(\omega,\pi), \quad det\, g=1.
\]
where $\zeta=\frac{\pi_1}{\pi_2}$, $h(x,\zeta)$ is regular in $|\zeta|>1-\epsilon$ and $k(x,\zeta)$ is regular in $|\zeta|<1+\epsilon$.
I adopt the Ansatz
\begin{eqnarray}
g_0&=&\left(\begin{array}{cc}e^{-\nu}&0\\
                           0&e^\nu\end{array}\right)
\left(\begin{array}{cc}\zeta^1&\rho\\
                       0&\zeta^{-1}\end{array}\right)
\left(\begin{array}{cc}e^{\mu}&0\\
                           0&e^{-\mu}\end{array}\right)\nonumber\\
&=&\left(\begin{array}{cc}e^\gamma\zeta^1&f(\gamma,\zeta)\\
                       0&e^{-\gamma}\zeta^{-1}\end{array}\right)
\end{eqnarray}

\begin{figure} [htb]
\includegraphics[width=6cm,angle=0,clip] {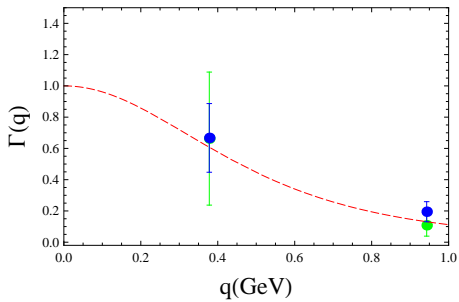} 
\caption{The proton charge form factor calculated with $16^3\times 32\times 16$ DWF configuration. Dashed line is the result of experiment.} \label{proton}
\end{figure}

In our 5-dimesional domain wall fermion case, $\gamma=\mu-\nu$ and $\mu, \nu$ contain the phase in the 5th direction $i\eta$. 
\begin{equation}
2\mu=i\omega_2/\pi_2-i\eta=(x_1+ix_2)\zeta+ix_0-x_3-i\eta \nonumber
\end{equation}
\begin{equation}
2\nu=i\omega_1/\pi_1+i\eta=(x_1-ix_2)\zeta+ix_0+x_3+i\eta \nonumber
\end{equation}
 The quaternion reality condition of the transformation matrix $g(\gamma,\zeta)$ gives
\begin{eqnarray}
&&\left(\begin{array}{cc}a_{L_s-1}& b_{L_s-1}\\
                       c_{L_s-1}& d_{L_s-1}\end{array}\right)
\left(\begin{array}{cc}\zeta^{1}e^\gamma& f\\
                       0&\zeta^{-1}e^{-\gamma}\end{array}\right)\nonumber\\
&&=\left(\begin{array}{cc}\zeta^{1}e^{-\gamma}& \bar f\\
                       0&\zeta^{-1}e^{\gamma}\end{array}\right)
\left(\begin{array}{cc}a_{0}& b_{0}\\
                       c_{0}& d_{0}\end{array}\right),
\end{eqnarray}
where $\displaystyle f=\frac{d_0 e^\gamma-a_0 e^{-\gamma}}{\psi}$, $\psi=c_{L_s-1}\zeta^1=c_0\zeta^{-1}$ 
and $\displaystyle \bar f=\overline{f(\bar\gamma,-\frac{1}{\bar\zeta})}$.
I search parameters using Mathematica and obtained Fig.\ref{proton}.

The charge form factor of DWF was calculated also in Schroedinger functional method \cite{LHPC09}, but the charge radius was smaller than the experiment. The difference of  the two is expected to be due to the treatment of final state interaction, as was the case in the $\eta$ decay into three mesons\cite {RT81}.

\subsection{The gluon self energy}
\'E. Cartan defined the vector field as a pl\"ucker coordinate of fermion spinors. 
The trilinear form of fermion, antifermion and vector field in the quaternion bases is
\begin{eqnarray}
&&{\mathcal F}=\phi^T CX\psi\nonumber\\
&&=x^1(\xi_{12}\xi_{314}-\xi_{31}\xi_{124}-\xi_{14}\xi_{123}+\xi_{1234}\xi_1)\nonumber\\
&&+x^2(\xi_{23}\xi_{124}-\xi_{12}\xi_{234}-\xi_{24}\xi_{123}+\xi_{1234}\xi_2)\nonumber\\
&&+x^3(\xi_{31}\xi_{234}-\xi_{23}\xi_{314}-\xi_{34}\xi_{123}+\xi_{1234}\xi_3)\nonumber\\
&&+x^4(-\xi_{14}\xi_{234}-\xi_{24}\xi_{314}-\xi_{34}\xi_{124}+\xi_{1234}\xi_4)\nonumber\\
&&+x^{1'}(-\xi_{0}\xi_{234}+\xi_{23}\xi_{4}-\xi_{24}\xi_{3}+\xi_{34}\xi_2)\nonumber\\
&&+x^{2'}(-\xi_{0}\xi_{314}+\xi_{31}\xi_{4}-\xi_{34}\xi_{1}+\xi_{14}\xi_3)\nonumber\\
&&+x^{3'}(-\xi_{0}\xi_{124}+\xi_{12}\xi_{4}-\xi_{14}\xi_{2}+\xi_{24}\xi_1)\nonumber\\
&&+x^{4'}(\xi_{0}\xi_{123}-\xi_{23}\xi_{1}-\xi_{31}\xi_{2}-\xi_{12}\xi_3)\end{eqnarray}

Using this trilinear form, I construct three loop gluon self-energy diagram as shown in Figs.2 and 3 %\ref{g11a} and  \ref{g22a} 
as transverse polarized and Figs.4 to 6 %\ref{g44a} to \ref{g44c} 
as Coulomb potential in the Coulomb gauge. Here the two exchanged vector fields are self-dual. 
The triality transformation transforms fermion field to other triality eigen states. If quark-gluon interaction is triality blind, the gluon created by a quark-anti quark pair in a triality sector will interact with quark-anti quark pairs of other triality. 
\begin{figure} [!b]\label{g11a}
\includegraphics[width=3cm,angle=0,clip] {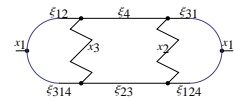}
 \caption{The self energy of tranverse polarizated gluon. There is another diagram with the chronological order reversed. }
\end{figure}
\begin{figure}[htb]\label{g22a}
\includegraphics[width=3cm,angle=0,clip] {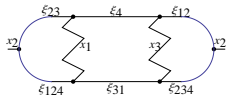}
  \caption{Same as the Fig.2. }
\end{figure}

\begin{figure} [htb]\label{g44a}
\includegraphics[width=3cm,angle=0,clip] {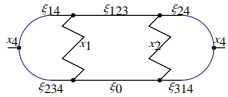}
\caption{The self energy diagram of the Coulomb potential. There is another diagram with the chronological order reversed.}
\end{figure}
\begin{figure}[htb]\label{g44b}
\includegraphics[width=3cm,angle=0,clip] {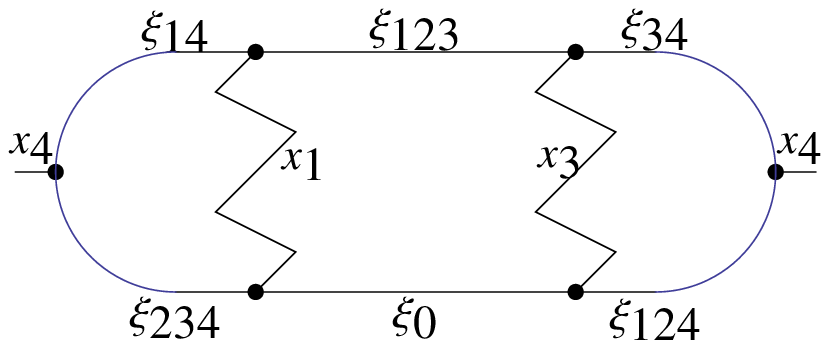}
\caption{Same as the Fig.4}
\end{figure}
\begin{figure}[htb]\label{g44c}
\includegraphics[width=3cm,angle=0,clip] {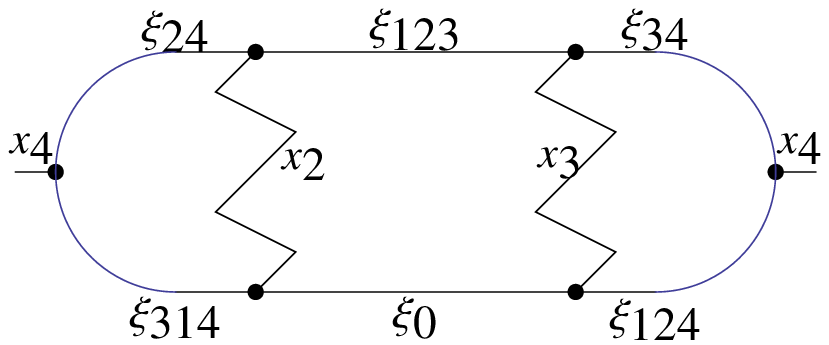}
\caption{Same as the Fig.4.}
\end{figure}

In finite temperature QCD, these diagrams give $g^6$ order term in the perturbative calculation of the pressure. Since all diagrams have the same phase, they are the candidate of compenating the $g^2$ order negative pressure term. When the conjecture of \cite{DD78} works, this kind of zero mode contribution dominates the plessure of the QCD ground state.

\subsection{Puzzles in the critical fermion number}
According to \cite{BZ82}, presence of infrared fixed point and the opening of the conformal window occurs in a region near a certain critical flavor number $N_f^c$. 
Lattice simulations in Schr\"odinger Functional(SF) method \cite{AFN08} says $N_f^c\sim 10$ while latttice simulation in MOM scheme \cite{SF10} and experimental data \cite{DBCK08} suggests presence of IR fixed point for $N_f^c=3$. 
In QCD there is axial anomaly, and to make the theory self-consistent, anomaly cancellation is required, but it is shown that it occurs when $N_f$ is larger than 10\cite{San09}. 

These puzzling features could be resolved, if the quark-gluon interaction is triality blind, consistent with the phenomenological analysis of weak decay processes \cite{Ma10}, and the effective $N_f$ in the SF scheme is three times larger than that in the MOM scheme.  

\section{Discussion and Conclusion}
In the review of Sh\"afer and Shyryak\cite{SS98}, the effective four quark interaction with auxiliary scalar field $L_a$ and $R_a$ is given as
\begin{eqnarray}
&&(\psi^\dagger\tau_a^-\gamma_-\psi)^2\to 2(\psi^\dagger\tau_a^-\gamma_-\psi)L_a-L_aL_a\nonumber\\
&&(\psi^\dagger\tau_a^-\gamma_+\psi)^2\to 2(\psi^\dagger\tau_a^-\gamma_+\psi)R_a-R_aR_a
\end{eqnarray}

In this model, meson decay into three mesons occur through exchange of self-dual gauge fields and/or quark-pair creation, but in $\eta_c\to \eta \pi\pi$ and $\eta_c\to K\bar K\pi$ decay processes, the exchange of two self-dual gauge fields  dominates. These decay could be measured in B-factory at KEK and yield useful information on instanton.

If charged lepton interaction preserves triality, but quark interactions does not, the hierarchy problem (fine tuning in the definition of GUT scale is necessary) and the $U(1)$ problem \cite{ChLi84} could be resolved.  Some unparticles \cite{Ge07}, which are believed to exist from astrophysical observations, could be quark-anti quark pairs that belong to different triality sectors from that of electrons or muons in the detector.

 Lattice simulations of larger lattice to confirm importance of the triality automorphism in IR QCD are under way.

%%%%%%%%%%%%%%%%%%%%%%%%%%%%%%%%%%%%%%%%%%%%%%%%
%% BACKMATTER
%%%%%%%%%%%%%%%%%%%%%%%%%%%%%%%%%%%%%%%%%%%%%%%%

%\begin{theacknowledgments}
\begin{center}
{\bf ACKNOWLEDGEMENTS}
\end{center}
The author thanks the organizer for the interesting conference in Madrid. Thanks are also due to computer centers at KEK and at Tsukuba Univertsity for supports.
%\end{theacknowledgments}

%%%%%%%%%%%%%%%%%%%%%%%%%%%%%%%%%%%%%%%%%%%%%%%%
%% The bibliography can be prepared using the BibTeX program or
%% manually.
%%
%% The code below assumes that BibTeX is used.  If the bibliography is
%% produced without BibTeX comment out the following lines and see the
%% aipguide.pdf for further information.
%%
%% For your convenience a manually coded example is appended
%% after the \end{document}
%%%%%%%%%%%%%%%%%%%%%%%%%%%%%%%%%%%%%%%%%%%%%%%%

%%%%%%%%%%%%%%%%%%%%%%%%%%%%%%%%%%%%%%%%%%%%%%%%
%% You may have to change the BibTeX style below, depending on your
%% setup or preferences.
%%
%%
%% For The AIP proceedings layouts use either
%%%%%%%%%%%%%%%%%%%%%%%%%%%%%%%%%%%%%%%%%%%%

%\bibliographystyle{aipproc}   % if natbib is available
\bibliographystyle{aipprocl} % if natbib is missing

%%%%%%%%%%%%%%%%%%%%%%%%%%%%%%%%%%%%%%%%%%%
%% You probably want to use your own bibtex database here
%%%%%%%%%%%%%%%%%%%%%%%%%%%%%%%%%%%%%%%%%%%
%\bibliography{sample}

%%%%%%%%%%%%%%%%%%%%%%%%%%%%%%%%%%%%%%%%%%%
%% Just a reminder that you may have to run bibtex
%% All of it up to \end{document} can be removed
%% if you don't like the warning.
%%%%%%%%%%%%%%%%%%%%%%%%%%%%%%%%%%%%%%%%%%%
%\IfFileExists{\jobname.bbl}{}
% {\typeout{}
%  \typeout{******************************************}
%  \typeout{** Please run "bibtex \jobname" to optain}
%  \typeout{** the bibliography and then re-run LaTeX}
%  \typeout{** twice to fix the references!}
%  \typeout{******************************************}
%  \typeout{}
% }

%\end{document}

%%%%%%%%%%%%%%%%%%%%%%%%%%%%%%%%%%%%%%%%%%%
%% The following lines show an example how to produce a bibliography
%% without the help of the BibTeX program. This could be used instead
%% of the above.
%%%%%%%%%%%%%%%%%%%%%%%%%%%%%%%%%%%%%%%%%%%

\end{document}